\documentclass[12pt]{amsart}
\usepackage{graphicx}
\usepackage{subfig}
\usepackage{tabularx}
\usepackage{tikz}
\usepackage{pdflscape}
\usepackage{graphicx}

\begin{document}
\newtheorem{proposition}{Proposition}
\newtheorem{corollary}{Corollary}
\newcommand{\ds}{\displaystyle}
\newtheorem{definition}{Definition}[section]
\newtheorem{theorem}{Theorem}
\newtheorem{lemma}{Lemma}
\newcommand\scaleSizeA{0.22}
\newcommand\scaleSizeB{0.14}
\parskip=12pt

%\ifpdf
%	\DeclareGraphicsExtensions{.pdf, .jpg, .tif}
%	\else
%	\DeclareGraphicsExtensions{.eps, .jpg}
%	\fi

\title[RRD Distributions]{Distributions of restricted rotation distances}

\author{Sean Cleary}
\address{Department of Mathematics \\
The City College of New York and the CUNY Graduate Center\\
City University of New York \\
New York, NY 10031}
\email{cleary@sci.ccny.cuny.edu}
\urladdr{{\tt http://cleary.ccnysites.cuny.edu}}

\author{Haris Nadeem}
\address{Department of Computer Science \\
The City College of New York\\
City University of New York \\
New York, NY 10031}
\email{haris.nadeem.bsc@gmail.com}

\thanks{
	This material is based upon work supported by the National Science Foundation under Grant No. \#1417820.
}

\begin{abstract}
Rotation distances measure the differences in structure between rooted ordered binary trees.
The one-dimensional skeleta of associahedra are rotation graphs, where two vertices representing trees are connected by an edge if they differ by a single rotation.
There are no known efficient algorithms to compute rotation distance between trees and thus distances in rotation graphs.
Limiting the allowed locations of where rotations are permitted gives rise to a number of notions of restricted rotation distances. Allowing rotations at a minimal such set of locations gives restricted rotation distance.  There are linear-time algorithms to compute restricted rotation distance, where there are only two permitted locations for rotations to occur.  The associated restricted rotation graph has an efficient distance algorithm.
There are linear upper and lower bounds on restricted rotation distance with respect to the sizes of the reduced tree pairs.  Here, we experimentally investigate the expected restricted rotation distance between two trees selected at random of increasing size and find that it lies typically in a narrow band well within the earlier proven linear upper and lower bounds.
\end{abstract}

\maketitle

\section{Introduction}

Binary trees capture hierarchical relationships in a wide range of settings.  For example, when there is an order on leaves, we have widely-used binary search trees, see Knuth \cite{knuth3}.  Simple local changes, called rotations, at nodes give rise to rotation distance and the the rotation graph, where two trees are connected by an edge in the rotation graph if they differ by a single rotation.  There are no known algorithms for computing rotation distance exactly in polynomial time, though there are some estimation algorithms which run in polynomial time of Baril and Pallo \cite{pallo} and Cleary and St.~John \cite{cleary2009linear} and the problem is known to be fixed-parameter tractable, see Cleary and St.~John \cite{rotfpt}.  
But there is thus no known algorithm for calculating distances efficiently in rotation graphs.
If we only allow rotations either all along the right arm of the tree or only at the root and right child of the root, then there are linear-time algorithms for computing the resulting restricted rotation distance and right-arm rotation distances, see Cleary \cite{rotipl} and Cleary and Taback \cite{rotbound}.
Thus we can explore the properties of distance in the related graph associated to
restricted rotation distance.
Here, we experimentally study the distributions of restricted rotation distance between randomly selected trees of increasing size and find that the distances appear to grow on average quite linearly with size with a linear coefficient of between three and four, with the distances distributed centrally arranged near the average in relatively narrow spreads.

This gives insight into the distribution of distances between pairs of trees in the restricted rotation graph which is not presently feasible for the rotation graph.

\section{Background and definitions}

In the following, by {\em tree} we mean a rooted binary tree where each node has either zero or two children, a left child and a right child.  Such trees are sometimes called {\em 0-2 trees} or {\em proper binary trees}. A node with no children is a {\em leaf}, and a node with two children is an {\em internal node}.   The {\em size} of a tree $T$ is the number of internal nodes in $T$. We number the $n+1$ leaves in a tree with $n$ internal nodes from left to right from $0$ to $n$.

\begin{figure}
    \centering
       
    \begin{tikzpicture}
\begin{scope} [node/.style={circle,draw},xshift=-0.25cm,scale=0.65,yshift=1.8cm]
   
	\node (r-1) at (0, 0) [node] {$$};
	\node (a-1) at (-1, -1) [node] {$$};
	\node (0-1) at (-2, -2) [node] {$$};
	\node (b-1) at (0, -2) [node] {$P$};
	\node (c-1) at (-1.5, -3) [node] {$$};
	\node (1-1) at (-2.5, -4) [node] {$$};
	\node (2-1) at (-0.5, -4) [node] {$$};
	\draw[-] (c-1) -- (1-1);
	\draw[-] (c-1) -- (2-1);
	\node (d-1) at (1.5, -3) [node] {$C$};
	\node (3-1) at (0.5, -4) [node] {$$};
	\node (e-1) at (2.5, -4) [node] {$$};
	\node (4-1) at (1.5, -5) [node] {$$};
	\node (5-1) at (3.5, -5) [node] {$$};
	\draw[-] (e-1) -- (4-1);
	\draw[-] (e-1) -- (5-1);
	\draw[-] (d-1) -- (3-1);
	\draw[-] (d-1) -- (e-1);
	\draw[-] (b-1) -- (c-1);
	\draw[-] (b-1) -- (d-1);
	\draw[-] (a-1) -- (0-1);
	\draw[-] (a-1) -- (b-1);
	\node (6-1) at (1, -1) [node] {$$};
	\draw[-] (r-1) -- (a-1);
	\draw[-] (r-1) -- (6-1);
\end{scope}
\end{tikzpicture}
\begin{tikzpicture}
\begin{scope} [node/.style={circle,draw},xshift=-0.25cm,scale=0.65,yshift=1.8cm]
  
	\node (r-1) at (0, 0) [node] {$ $};
	\node (a-1) at (-1, -1) [node] {$ $};
	\node (0-1) at (-2, -2) [node] {$ $};
	\node (b-1) at (0, -2) [node] {$C'$};
	\node (c-1) at (-2.5, -4) [node] {$ $};
	\node (1-1) at (-3.5, -5) [node] {$ $};
	\node (2-1) at (-1.5, -5) [node] {$ $};
	\draw[-] (c-1) -- (1-1);
	\draw[-] (c-1) -- (2-1);
	\node (d-1) at (-1.5, -3) [node] {$P'$};
	\node (3-1) at (-0.5, -4) [node] {$ $};
	\node (e-1) at (1.5, -3) [node] {$ $};
	\node (4-1) at (.5, -4) [node] {$ $};
	\node (5-1) at (2.5, -4) [node] {$ $};
	\draw[-] (e-1) -- (4-1);
	\draw[-] (e-1) -- (5-1);
	\draw[-] (d-1) -- (3-1); 
	\draw[-] (d-1) -- (c-1);
	\draw[-] (b-1) -- (e-1);
	\draw[-] (b-1) -- (d-1);
	\draw[-] (a-1) -- (0-1);
	\draw[-] (a-1) -- (b-1);
	\node (6-1) at (1, -1) [node] {$$};
	\draw[-] (r-1) -- (a-1);
	\draw[-] (r-1) -- (6-1);
\end{scope}

\end{tikzpicture}
  \caption{An example of a left rotation at node $P$, with a rotation promoting child node $C$ to $C'$ and demoting parent node $P$ to $P'$.  The left hand tree has encoding $1101100101000$ and the right hand tree is $1101110001000$.  All other nodes are unaffected by the rotation at $P$.  Right rotation at $C'$ is the inverse operation, taking the tree on the right to the tree on the left.}
    \label{rotfig}
\end{figure}
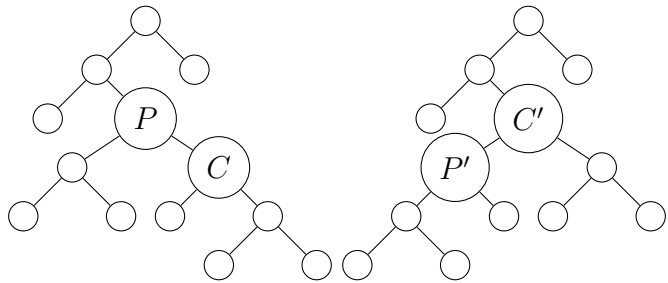

 We encode binary trees via the standard encoding of a preorder traversal where an internal node is denoted by 1 and a leaf node by 0.  So the left hand tree in Figure \ref{rotfig} has encoding 1101100101000 and the right hand tree has encoding 1101110001000. A {\em rotation} at a node $P$ is the operation depicted in Figure \ref{rotfig} where one grandchild of $P$ is promoted to become a child of $P$, one child is demoted to become a grandchild, and where one grandchild's parent node is switched in an order-preserving way.  In terms of encodings, a left rotation at a node can be regarded as a string substitution of the form $\ldots 1x1yz \ldots$ becoming $ \ldots 11xyz \ldots$  where $x,y$, and $z$ are encodings of subtrees, with a right rotation the inverse string substitution operation.

Given two trees $S$ and $T$ of size $n$, Culik and Wood \cite{culik1982note} showed that there is always at least one sequence of rotations transforming $S$ to $T$ and thus defined rotation distance. 
 {\em Rotation distance} between $S$ and $T$, denoted $d(S,T)$, is the  minimum number of rotations needed to transform $S$ to $T$ where the rotations are permitted at any nodes present.   
 We need not have rotations permitted at every node to transform any tree to any other- a minimal set of permitted rotations has size 2, as described by Cleary \cite{rotipl}.  We take those two locations to be the root and the right child of the root, giving  {\em restricted rotation distance} between $S$ and $T$, denoted $d_R(S,T)$, as the  minimum number of rotations needed to transform $S$ to $T$ where the rotations are permitted only at the root node and the right child of the root node, if present.  
 
 The {\em rotation graph $RG(n)$} of size $n$ is the graph whose vertices are rooted binary trees of size $n$ and where two vertices are connected by an undirected edge if there is a single rotation transforming the one tree to the other.  The rotation graph is the one-dimensional skeleton of the associahedron of the appropriate size.  The notions foundational to the geometric realization of associahedra go back to Tamari \cite{tamari} and Stasheff \cite{stasheff} and were first published concretely by Lee \cite{lee}.  Here, we consider
distances in the related {\em restricted rotation graph $RRG(n)$} where the vertices are again trees and an edge is present between trees $S$ and $T$ if they differ by a single rotation at either the root or the right child of the root. 

 A {\em tree pair} (S,T) is a pair of trees of the same size.
A tree pair $(S,T)$ is {\em unreduced} if there are nodes in both $S$ and $T$ such that leaf node children numbered as $i$ and $i+1$, via preorder traversal of the tree, are the same in both trees.  A {\em reduction} in a tree pair is the removal of such a pair of identically numbered siblings in each tree, replacing them with a single leaf $i$,  and then renumbering to get a new tree pair $(S',T')$ of one smaller size.
A tree pair $(S,T)$ is said to be {\em reduced} if there are no possible reductions.  Note that for both rotation distance and restricted rotation distances, the distances between $S$ and $T$ are the same as between the representatives of their reduced tree pair $S'$ and $T'$ as the same sequence of rotations will perform the required transformations, see \cite{rotipl}.
The {\em binary address} of a node in a tree is a sequences of 0's and 1's representing the path from the root to the node with a 0 for each left child and 1 for each right child.  For example, the address of node $C$ in Figure \ref{rotfig} in the left hand tree is $011$ as the path from the root to $C$ is a left edge followed by two right edges.

\begin{figure}
    \centering

\begin{tikzpicture}
\begin{scope} [node/.style={circle,draw},xshift=-0.25cm,scale=0.85,yshift=1.8cm]
  
	\node (r-1) at (0, 0) [node] {$5$};
	\node (a-1) at (-1, -1) [node] {$0$};
	\node (0-1) at (-2, -2) [node] {\color{red} 0};
	\node (b-1) at (0, -2) [node] {$3$};
	\node (c-1) at (-2.5, -4) [node] {$1$};
	\node (1-1) at (-3.5, -5) [node] {\color{red} 1};
	\node (2-1) at (-1.5, -5) [node] {\color{red} 2};
	\draw[-] (c-1) -- (1-1);
	\draw[-] (c-1) -- (2-1);
	\node (d-1) at (-1.5, -3) [node] {$2$};
	\node (3-1) at (-0.5, -4) [node] {\color{red} 3};
	\node (e-1) at (1.5, -3) [node] {$4$};
	\node (4-1) at (.5, -4) [node] {\color{red} 4};
	\node (5-1) at (2.5, -4) [node] {\color{red} 5};
	\draw[-] (e-1) -- (4-1);
	\draw[-] (e-1) -- (5-1);
	\draw[-] (d-1) -- (3-1); 
	\draw[-] (d-1) -- (c-1);
	\draw[-] (b-1) -- (e-1);
	\draw[-] (b-1) -- (d-1);
	\draw[-] (a-1) -- (0-1);
	\draw[-] (a-1) -- (b-1);
	\node (6-1) at (1, -1) [node] {\color{red} 6};
	\draw[-] (r-1) -- (a-1);
	\draw[-] (r-1) -- (6-1);
\end{scope}

\end{tikzpicture}
  \caption{A tree of size 6 with leaves numbered in red from 0 to 6 and with internal nodes numbered from 0 to 5.  Nodes 0 and 5 are left nodes and all other internal nodes are interior nodes. }
    \label{nodefig}
\end{figure}
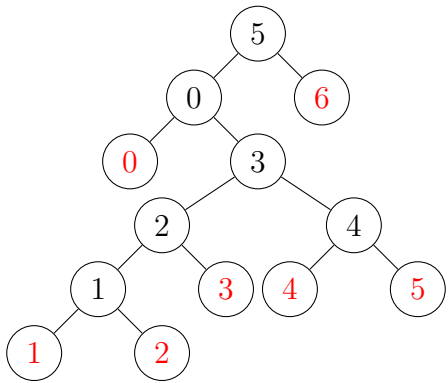

A  {\em right node} of a tree is one whose binary address consists only of 1's and has at least one 1. A {\em left node} is one whose binary address consists only of 0's.  The root node is thus a left node but not a right node.  All non-right and non-left nodes of a tree are 
{\em interior nodes}.  We number nodes with an in-order traversal of the tree, and a
{\em node pair} from a tree pair $(S,T)$ is a pair of nodes numbered the same in such traversals.  Figure \ref{nodefig} shows leaves and nodes numbered in the resulting left-to-right in-order traversals of leaves and interior nodes respectively.

To calculate restricted rotation distance, we use the methods of Fordham \cite{blakegd}.  His methods were designed to calculate word length exactly in Thompson's group $F$ with respect to the generating set  $\{x_0,x_0^{-1},x_1,x_1^{-1}\}$, and give a minimal length representative of a word with respect to that generating set.  The generator $x_0$ corresponds to right rotation at the root, with $x_0^{-1}$ correspondingly the inverse which is a left rotation at the root.  Similarly, $x_1$ and its inverse correspond to rotations at the right child of the root.  So word length in $F$ translates into restricted rotation distance between trees, as described in \cite{rotipl, rotbound}.  

His method takes as input two trees forming a reduced tree pair, and classifies each interior node as one of seven types as follows:
\begin{itemize}

\item
$L_0$: The first node on the left side of the tree.
\item
$L_L$:  Any left node other than the leftmost node.

\item
$I_0$: An interior node with no right child.

\item
$I_R$: An interior node with a right child.

\item
$R_I$: Any right node numbered $k$ whose immediate successor node $k+1$ is an interior node.

\item
$R_{NI}$: A right node which is not of type $R_I$ but for which there is some successor interior node.

\item
$R_0$: A right node with no successor interior node.
\end{itemize}

A primary result of  Fordham \cite{blakegd} is that the word length $|w|$ in Thompson's group $F$ with respect to the standard finite generating set can be calculated by classifying node pairs into those seven types and summing the
totals from the table below.  Note that the first node pair is always of type $(L_0,L_0)$ and adds weight 0,  and the single $L_0$ in each tree must necessarily be paired, so  $L_0$ is not listed Table \ref{caretweights}.

\begin{table}
\begin{center}
\begin{tabular}{|c|c|c|c|c|c|c|}

\hline
 & $R_0$ & $R_{NI}$ & $R_I$ & $L_l$ & $I_0$ & $I_R$ \\
 \hline

 $R_0$ & 0 & 2 & 2 & 1 & 1 & 3 \\ \hline
 $R_{NI}$ & 2 & 2 & 2 & 1 & 1 & 3 \\ \hline
 $R_I$ & 2 & 2 & 2 & 1 & 3 & 3 \\ \hline
 $L_l$ & 1 & 1 & 1 & 2 & 2 & 2 \\ \hline
 $I_0$ & 1 & 1 & 3 & 2 & 2 & 4 \\ \hline
$I_R$ & 3 & 3 & 3 & 2 & 4 & 4 \\ \hline
\end{tabular}

\end{center}
\caption{Weights for caret pairs by caret pair types.}\label{caretweights}.
\end{table}

As described \cite{rotbound}, since all non-$L_0$ carets contribute at least one to word length (and thus at least one to restricted rotation distance), and since a caret can contribute at most $4$ to word length, analysis of caret types and configurations give that the restricted rotation distance between two trees of size $n$ lies between  $n-1$ and $4n-8$ and is sharp for $n \geq 3$.  Fordham's method goes further and can be in fact used to not only find restricted rotation distances, but also to find and enumerate all possible minimal length paths between the relevant trees.  We note that there have been computations to calculate the number of words of Thompson's group $F$ of increasing word length with respect to the standard generating set (and thus restricted rotation distances) of increasing sizes by Burillo, Cleary, and Weist \cite {fexplore} and Elder, Fusy, and Rechnitzer \cite{eldercountf}, with the latter giving the first 1500 terms of the OEIS sequence A156945 \cite{oeisthomp} which are the number of elements of increasing word length size.  The relationship between word length size and tree size is linear but knowing word length gives only linear bounds on the tree size.

\section{Distributions of restricted rotation distance}

We study computationally the distribution of restricted rotation distance between rooted binary trees.  Work of Cleary and Maio \cite{rotdist} analyzes distributions of ordinary rotation distances.  Here, we address similar questions for restricted rotation distances.  The general question is: given two trees of the same size $n$, what is the expected restricted rotation distance between them?  We anticipate that on average, larger tree pairs have larger distances between them, but we would like to estimate the rates of growth as well as the dispersal.  Work of Cleary and Taback \cite{rotbound} gave sharp lower and asymptotically sharp upper bounds for restricted rotation distances, and we find that the vast majority of instances are clustered centrally and not near the bounds.

\begin{table}
\begin{tabular}{|c|c|c|c|} \hline
Tree size range &  \# sampled & Avg. red. frac. & Avg. RRD ratio \\ \hline
10--19 & 138999 & 0.907533 & 2.24473 \\ 

20--29 & 161500 & 0.917172 & 2.64333 \\ 

30--39 & 150500 & 0.920593 & 2.83326 \\ 

40--49 & 133000 & 0.922663 & 2.9421 \\ 

50--59 & 144000 & 0.923896 & 3.00793 \\ 

60--69 & 134500 & 0.924459 & 3.05513 \\ 

70--79 & 129000 & 0.924884 & 3.08993 \\ 

80--89 & 119000 & 0.925221 & 3.1151 \\ 

90--99 & 118500 & 0.925659 & 3.13679 \\ 

100--199 & 685191 & 0.92646 & 3.19676 \\ 

200--299 & 509390 & 0.927268 & 3.24813 \\ 

300--399 & 310962 & 0.927496 & 3.26887 \\ 

400--499 & 111460 & 0.927678 & 3.27999 \\ 

500--599 & 89580 & 0.92783 & 3.28727 \\ 

600--699 & 100600 & 0.927795 & 3.29198 \\ 

700--799 & 102600 & 0.9279 & 3.29606 \\ 

800--899 & 43600 & 0.927921 & 3.29866 \\ 

900--999 & 45450 & 0.928027 & 3.30121 \\ 

1000--1249 & 89200 & 0.928008 & 3.30416 \\ 

1250--1499 & 86000 & 0.928002 & 3.3069 \\ 

1500--1749 & 99000 & 0.928071 & 3.30908 \\ 

1750--1999 & 35600 & 0.928121 & 3.31039 \\ 

2000--2249 & 20000 & 0.928089 & 3.31145 \\ 

2250--2499 & 19800 & 0.928089 & 3.31235 \\ 

2500--2749 & 18764 & 0.928117 & 3.31311 \\ 

2750--2999 & 13900 & 0.928124 & 3.31386 \\ 

3000--3249 & 12124 & 0.928094 & 3.31407 \\ 

3250--3499 & 8044 & 0.928185 & 3.31517 \\ 

3500--3999 & 3072 & 0.928024 & 3.31562 \\ 

4000--4500 & 800 & 0.928023 & 3.31568 \\  \hline
\end{tabular}  
\caption{Tree pair restricted rotation distances for unreduced tree pairs.  Given are the average fractions of the reduced tree pairs size of the originally generated tree pair size and the average ratio of restricted rotation distance to the generated tree pair size.}
\label{RRDfull}
\end{table}

\begin{figure}
  \centering
    \includegraphics[width=0.80\textwidth]{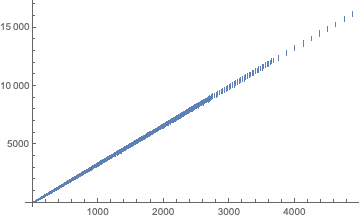}
  \caption{Restricted rotation distance vs. raw size for randomly selected tree pairs of increasing sizes.}
  \label{fig:rrdvraw}
\end{figure}

\begin{table}
\begin{tabular}{|c|c|c|} \hline
Tree size range  &  Number of tree pairs sampled &  Average RRD size  \\ \hline
10--19 & 168846 & 2.609 \\ 
20--29 & 166650 & 2.96244 \\ 
30--39 & 145364 & 3.12548 \\ 
40--49 & 152971 & 3.22228 \\ 
50--59 & 144317 & 3.28264 \\ 
60--69 & 139509 & 3.32627 \\ 
70--79 & 132652 & 3.35818 \\ 
80--89 & 126454 & 3.38269 \\ 
90--99 & 94370 & 3.40162 \\ 
100--199 & 700470 & 3.45925 \\ 
200--299 & 504029 & 3.50732 \\ 
300--399 & 272408 & 3.52717 \\ 
400--499 & 116513 & 3.53867 \\ 
500--599 & 97243 & 3.54577 \\ 
600--699 & 107923 & 3.55041 \\ 
700--799 & 74740 & 3.55356 \\ 
800--899 & 48099 & 3.55662 \\ 
900--999 & 40737 & 3.55859 \\ 
1000--1249 & 94865 & 3.56172 \\ 
1250--1499 & 100074 & 3.56451 \\ 
1500--1749 & 77950 & 3.56616 \\ 
1750--1999 & 22109 & 3.56769 \\ 
2000--2249 & 21630 & 3.56872 \\ 
2250--2499 & 20622 & 3.5695 \\ 
2500--2749 & 15851 & 3.57045 \\ 
2750--2999 & 13158 & 3.57073 \\ 
3000--3249 & 8342 & 3.57162 \\ 
3250--3499 & 2607 & 3.57268 \\ 
3500--3999 & 821 & 3.57276 \\ 
4000--4500 & 717 & 3.57285 \\  \hline
\end{tabular}  
\caption{Tree pair restricted rotation distances divided by tree pair size, for reduced tree pairs of increasing size ranges.}
\label{RRDred}
\end{table}

\begin{figure}
  \centering
    \includegraphics[width=0.80\textwidth]{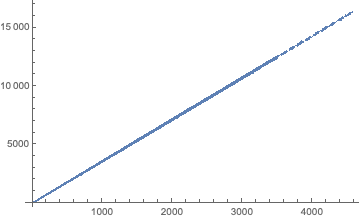}
  \caption{Restricted rotation distance vs. reduced size for randomly selected tree pairs of increasing sizes, by the size of the resulting reduced tree pair after reduction.}
  \label{fig:rrdvred}
\end{figure}

We sample rooted binary tree pairs at random using Remy's algorithm \cite{remy} for each tree,  which guarantees a uniform randomly generated tree of size $n$.
Work on the asymptotic density of isomporphism classses of subgroups of Thompson's group $F$ of Cleary, Elder, Rechnitzer and Taback \cite{randomf} addresses the question of the expected fraction of tree pairs which are reduced, and later work of Cleary, Rechnitzer and Wong \cite{commonedges} describes the asymptotics of the expected sizes of reduced components of tree pairs. 

Here, we study two main questions: 
\begin{itemize}
\item Given two trees selected at random of size $n$, what is the expected restricted rotation distance between them?
\item Given a reduced tree pair of size $n$, what  is the expected restricted rotation distance between the pair?
\end{itemize}

We generated trees pairs $(S,T)$ at random, then calculated the reduced representatives $(S',T')$ of each tree pair, then the corresponding restricted rotation distance, $d_R(S,T)=d_R(S',T')$, which are the same as the reductions reflect commonality which does not change the distance.

We note that generating reduced tree pairs of a specified size is not as feasible as generating tree pairs generally.    As described in \cite{commonedges} and \cite{randomf}, a tree pair selected at random is likely to have a number of reductions, and the resulting reduced representative is on average about 10\% smaller.  But of course there is a (increasingly small) chance that the generated tree is already reduced, and also a (vanishingly small) chance that it reduces all the way down to the empty tree pair.  Cleary, Rechnitzer and Wong \cite{commonedges} analyze some properties of the distribution of the resulting sizes of reduced tree pairs.  Cleary and Maio \cite{hardcases} have an algorithm which guarantees to produce not only a reduced tree pair of a specified size, but is difficult in an additional sense as well-- not having any obvious initial first moves along minimal length paths.  Unfortunately, that algorithm does not choose uniformly from among the possible ones.  The particular number of such difficult instances is not even known precisely, though  Cleary and Maio \cite{hardcounts} calculate the number of such cases exhaustively for small sizes and approximately for larger ones.

By generating large families of trees across a range of sizes and then performing reductions, we get a range of reduced tree pairs to consider and analyze.  The resulting reduced tree pairs are necessarily smaller than the generated, possibly reducible, tree pairs, but since the number of reductions vary, there is a dispersal in the resulting sizes of the reduced tree pairs.  That is, if we generate 1400 tree pairs of size 1000, the smallest resulting reduced pair may be  896 and the largest 955, with a mean and median of about 928 with the most commonly occurring being  929 with 73 occurrences.  The tree pairs were generated of fixed sizes, often 500 apart.  Thus, after reductions, these sizes would reduce to different extents which may lead to gaps in the resulting reduced sizes.  So we generate many examples across a range of increasing sizes in an effort to get representative samples across a broad range.

\section{Experiments and Discussion}

For the computational experiments we described, we generated about 3.6 million tree pairs of sizes ranging from 10 to 4400.  We  reduced each tree pair to a reduced representative, and then calculated the restricted rotation distances using Fordham's method.

To compare average restricted rotation distances across a range of sizes, we consider the {\em RRD ratio}, which for a tree pair $(S,T)$ of size $n$ is $d_R(S,T)/n$.  This gives a somewhat normalized measure of the typical contribution of tree carets to the restricted rotation distance and a sense of how quickly the restricted rotation distance grows with increased tree size.  We note that trees realizing the the lower bound of restricted rotation distance from \cite{rotbound} would have an RRD ratio limiting to 1, and those realizing the the upper bound would have an RRD ratio limiting to 4.

Table \ref{RRDfull} tabulates the results across a range of unreduced sizes, with Figure \ref{fig:rrdvraw} plotting the results for these unreduced sampled tree pairs.  We see tight linear behavior of distance with respect to raw size, despite the fact that the amount of reductions vary considerably and the resulting sizes have a large influence on the corresponding distances.

Owing to the time of computation, larger size tree pairs were not sampled as extensively as the smaller ones.
 In Figure \ref{fig:rrdvraw} the sampling increments of size 500 are visible, and in Figure \ref{fig:rrdvred} the fact that those sizes have dispersed somewhat as the reductions in size vary is visible.  The fraction of common edges in a more general sense was computed asymptotically by Cleary, Rechnitzer and Wong \cite{commonedges} to be $6-\frac{16}{\pi} \sim 0.907$, so the observed fractions of reduced size from generated size of about 0.928 is consistent with that.  That asymptotic analysis allowed reductions of internal common edges in addition to the peripheral ones relevant to the tree reductions considered here.

In the remaining analyses, we restrict our attention to the resulting generated reduced tree pairs as the distances are more tightly related to the sizes after reduction.

Table \ref{RRDred} tabulates the distances observed across a range of reduced tree pair sizes, and  Figure \ref{fig:rrdvred} plots these results.  We can again see tight linear behavior, where the reduced trees have on average larger rotation distances and a smaller spread in the observed reduced instances relative to the unreduced sizes.

The examples from Cleary and Taback \cite{rotbound} giving the bounds of $n-1 \leq d_R(S,T) \leq 4n-8$ are clearly quite constrained, as the vast majority of the sampled lengths lie close to about $3.57n$, well away from the upper and lower bounds.  
We note that in both cases, the maximum possible distances (about 4 times the size) and minimal possible distances (one less than the size) lie far away from the randomly-generated instances.
This is not surprising as those examples to show the sharpness of the bounds were carefully constructed in a very specific manner to realize those bounds.  

We note that the only entries in Table \ref{caretweights} that contribute 4 to restricted rotation distance are $(I_R, I_0)$ and $(I_R, I_R)$ which involve interior carets being paired with interior carets.  Given that the average distances are well above 3, such caret pairings are necessarily quite common and cannot occur in the examples realizing the lower bounds of $n-1$.

\begin{figure}
  \centering
    \includegraphics[width=0.80\textwidth]{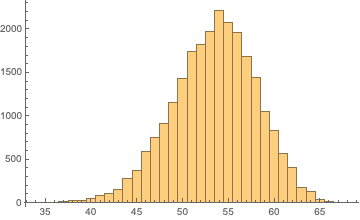}
  \caption{Distribution of restricted rotation distances for 24,067 randomly-produced reduced tree pairs of size 19.
  The sample mean is about 53.5
  and the sample standard deviation is about 4.58.
  }
  \label{fig:hist19}
\end{figure}

\begin{figure}
  \centering
    \includegraphics[width=0.80\textwidth]{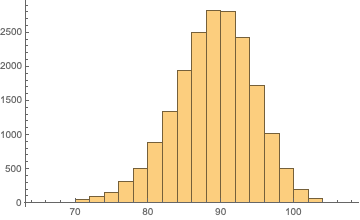}
  \caption{Distribution of restricted rotation distances for 19,307 randomly-produced reduced tree pairs of size 29.
   The sample mean is about 88.5
     and the sample standard deviation is about  5.45.
}
  \label{fig:hist29}
\end{figure}
\begin{figure}
  \centering
    \includegraphics[width=0.80\textwidth]{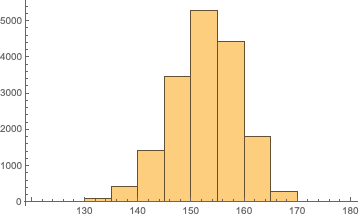}
  \caption{Distribution of restricted rotation distances for 17,196 randomly-produced reduced tree pairs of size 47.
   The sample mean is about 152.3
     and the sample standard deviation is about  6.36.
}
  \label{fig:hist47}
\end{figure}
\begin{figure}
  \centering
    \includegraphics[width=0.80\textwidth]{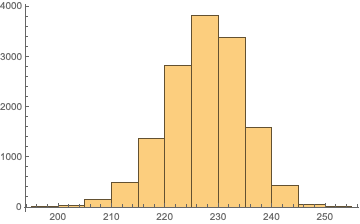}
  \caption{Distribution of restricted rotation distances for 14,155 randomly-produced reduced tree pairs of size 68.
   The sample mean is about 227.1
     and the sample standard deviation is about  7.20.
}
  \label{fig:hist68}
\end{figure}
\begin{figure}
  \centering
    \includegraphics[width=0.80\textwidth]{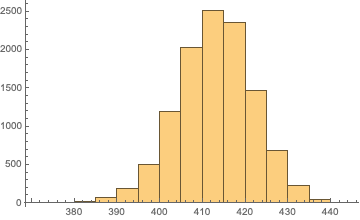}
  \caption{Distribution of restricted rotation distances for 11,258 randomly-produced reduced tree pairs of size 120.
   The sample mean is about 412.6
     and the sample standard deviation is about  8.79.
}
  \label{fig:hist120}
\end{figure}
\begin{figure}
  \centering
    \includegraphics[width=0.80\textwidth]{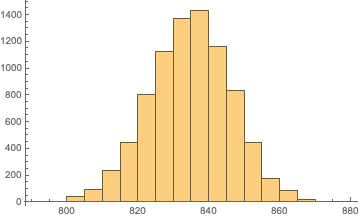}
  \caption{Distribution of restricted rotation distances for 8266 randomly-produced reduced tree pairs of size 238.
   The sample mean is about 834.3
     and the sample standard deviation is about  11.4.
}
  \label{fig:hist238}
\end{figure}

\begin{figure}
  \centering
    \includegraphics[width=0.80\textwidth]{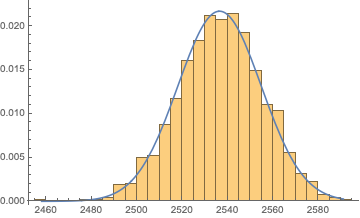}
  \caption{Distribution of restricted rotation distances for 1200 randomly-produced reduced tree pairs of size 714.
   The sample mean is about 2536.4.
     and the sample standard deviation is about  18.4.  A normal distribution with the same mean and standard deviation is superimposed for comparison.
}
  \label{fig:hist714}
\end{figure}

Not surprisingly, given the strong linear behavior observed, a fitted linear model agrees with the sampled data exceptionally well, giving 
$d_R(S,T) \sim  3.31941 n -17.0321 $ for restricted rotation distance in terms of https://www.overleaf.com/project/594bf4c7597a6dfa7c7cd2cb unreduced tree pair sizes $n$, and
$d_R(S,T) \sim   3.57612 n -16.1551$ correspondingly for reduced tree pairs of size $n$.

  We see that the standard deviations of the observed RRD ratios of restricted rotation distance are relatively small and stable, dropping steadily from about 0.33 for the smallest size trees sampled, to about 0.025 for tree sizes in the hundreds, then dropping to about 0.01 for tree sizes in the hundreds, with an observed average standard deviation of ratios of 0.009 for the largest tree sizes sampled.  These are for the normalized ratios- the standard deviations do increase with size, albeit somewhat more slowly.

The distributions of restricted rotation for reduced tree pairs of a fixed size show an approximately normal shape, slightly skewed to the left for smaller sizes but less so for larger sizes.  Here, we chose a few sizes for which there were a reasonable number of observed instances, shown in Figures \ref{fig:hist19} to Figure \ref{fig:hist714}.  These distributions have characteristic normal shapes, and further suggest that the extremely short and extremely long cases shown earlier to be possible are exceptionally rare. The vast majority of randomly-selected cases lie in relatively narrow bands concentrated on a line well away from the lowest and highest possible bounds.  For the largest million tree pairs sampled, less than 175,000 were more than 1\% away from the distance predicted by the linear model, and all but 1054 were within 3\% of the linear prediction, with the largest observed deviation from the linearly fitted model being less than 5\% away from the predicted distance.

Thus we have developed some understanding of typical behavior of distances in restricted rotation graphs $RRG(n)$ for a decent range of distances.  Note that analyzing the corresponding questions for the rotation graphs  $RG(n)$ are not presently feasible beyond size about 20 even experimentally due to the difficulty of computing ordinary rotation distance exactly., where the best known algorithms have exponential running time in the size of the trees.

\bibliographystyle{plain}

\end{document}